\documentstyle [12pt] {article}

\def\L{{\cal L}}
\def\R{{\rm I \!\!\, R}}
\def\be{\begin{equation}}
\def\ee{\end{equation}}
\def\bea{\begin{eqnarray}}
\def\eea{\end{eqnarray}}

\begin{document}
\baselineskip=12pt

\title{ {\bf Kinematic self-similar locally rotationally symmetric
 models }}

\author{ A.M. Sintes\thanks{{\sl Email:} sintes@aei-potsdam.mpg.de}\\
 {\small Max-Planck-Institut f\"ur Gravitationsphysik
 (Albert-Einstein-Institut),}\\
{\small Schlaatzweg 1, D-14473 Potsdam. Germany}}
\date{}
\maketitle

\begin{abstract}
A brief summary of results on kinematic self-similarities in general relativity
is given. Attention is  focussed on
locally rotationally symmetric models  admitting
kinematic self-similar vectors.
Coordinate expressions for the metric and the kinematic self-similar vector are
provided.
 Einstein's field equations for perfect fluid models are 
investigated and {\it all} the homothetic  perfect fluid solutions
admitting a maximal four-parameter group of isometries are given.
\end{abstract}
{\bf PACS numbers:} 02.40.K, 04.20.Jb, 04.40.Nr, 98.80.Hw

\section{Introduction}
This article is focused on kinematic self-similar models exhibiting a 
maximal four-parameter  group of isometries, $G_4$ (in addition to the
self-similar symmetry). Only perfect fluid models are considered.

Perfect fluid solutions admitting a maximal simply-transitive
group $G_4$ are all known. They correspond to the homogeneous
Ozsv\'ath solutions \cite{O}. In the multiply-transitive case,
they are locally rotationally symmetric  (LRS), 
the maximal group $G_4$ acts on three-dimensional
non-null orbits $S_3$ or $T_3$,
and the solutions are algebraically special. 
Such models are the concern of
this paper. Their metrics can be written in the forms  
\cite{Ellis67,Stewart68}
\bea
{\rm (i)} &\,& ds^2=\epsilon
(dt^2-A^2(t)dx^2)+B^2(t)(dy^2+\Sigma^2(y,k)dz^2) \ ,\label{r8}\\
{\rm (ii)} &\,& ds^2=\epsilon
(dt^2-A^2(t)\sigma^2)+B^2(t)(dy^2+\Sigma^2(y,k)dz^2) \ ,\label{r9}\\
{\rm (iii)} &\,& ds^2=\epsilon (dt^2-A^2(t)dx^2)+B^2(t)e^{2x}(dy^2+dz^2)
\ ,\label{r10}
\eea
with $\epsilon=\pm 1$, $k=0,\pm 1$ and $\sigma=dx+\Gamma(y,k)dz$, where
\be
\label{Sigma}
\Gamma(y,k)=\left\{ \begin{array}{ll}
\cos y  & k=+1\\
y^2  & k=0\\
\cosh y & k=-1
\end{array}
\right.
\ , \quad
\Sigma(y,k)=\left\{ \begin{array}{ll}
\sin y & k =+1 \\
y & k =0 \\
\sinh y & k =-1\ .
\end{array}
\right.
\ee

The known exact solutions of the metric (i) with ${\epsilon=-1}$ have been
 collected
by Vajk and Eltgroth \cite{Vajk}, and 
the field equations for the metric (iii) have been qualitatively studied by
Collins \cite{Collins74} and Shikin \cite{Shikin75}.
 General LRS space-times have been investigated many times in the literature
\cite{Kramer}. The symmetry groups of the LRS  models have been
discussed by Stewart and Ellis \cite{Stewart68}, van Elst and Ellis \cite{E},
van Elst and Uggla \cite{EU}.
 More recently, LRS perfect fluids have been studied by Marklund \cite{M},
 and Nilsson and Uggla \cite{NU}.
 
 In this paper, we study LRS perfect fluids (with a maximum four-parameter
 group of isometries) admitting kinematic self-similar vectors.

The concept of kinematic self-similarity was first introduced by Carter and 
Henriksen \cite{c1,c2}  as  a generalization to the homothety \cite{CaT}
and as a more natural counterpart of the concept of self-similarity present
 in Newtonian 
mechanics.

A vector field $X$  is called a kinematic self-similar vector
field (KSS) if it satisfies the conditions \cite{c1,c2}
\be
\L_X u_a=\alpha u_a \ , \qquad \L_X h_{ab}= 2 \delta h_{ab} \ , \label{k1}
\ee
where $\alpha$  and $\delta$ are constants, $\L$ stands for the Lie 
derivative operator, $u^a$ is the four-velocity of the fluid, and
$h_{ab}=g_{ab}+u_au_b$
is the projection tensor  of the metric
into the three-spaces orthogonal to $u^a$. 

The kinematic self-similar transformations 
are characterized in a well defining way
by the scaling-independent ratio $\alpha /\delta$ which is referred 
to as the similarity index. This index is finite except in the case
 of rigid transformations which is characterized by $\delta=0$ and
referred to as type infinite.
In the case $\alpha
=\delta$, it follows that $X$ is a homothetic vector field (HVF) 
\cite{CaT}   and, evidently, if $\alpha = \delta =0$, then
 $X$ becomes a Killing vector
(KV).
 Another case of special interest is  the
type zero  (i.e., $\alpha=0$) which corresponds to space dilatation 
without time amplification.

The study  of self-similar models  is very important in general relativity.
Some self-similar models are tractable because their symmetry makes them
less complicated. Besides their intrinsic mathematical interest,
self-similar solutions play an important role
 in   astrophysics and cosmology 
(see \cite{Coley,Carr} for a detail review).
The astrophysical applications include gravitational collapse and the
occurrence of naked singularities. The cosmological applications include
features of gravitational clustering, cosmic voids, the formation of
bubbles at a cosmological phase-transition in the early universe,
explosions in an (expanding) homogeneous background, 
cosmological models containing black holes, and their possible role
as asymptotic states for more general models.

The rest of the
 paper is organized as follows. The next section contains a brief 
summary of results on kinematic self-similar vector fields
 and space-times admitting
them.
 In section 3, we analyze
 LRS models constrained to admit KSS and we provide coordinate expressions
 for the metric and the KSS.
In
section 4,
we present the different perfect fluid solutions. The study is exhausted
for the metrics  (i) and (ii). For the metric (iii), we distinguish
the cases where  the fluid flow is comoving from where it is non-comoving.
In the comoving case, the only possible perfect fluid solution is a 
special case of a Friedmann-Robertson-Walker
 (FRW) model and,  in the non-comoving case, we show that
 perfect fluid solutions are non-homothetic.
Moreover, since the case of simply-transitive group $G_4$ is empty of
homothetic solutions \cite{yo,Hall90}, we explicitely give {\it all} the 
homothetic  solutions
admitting a maximal 4-dimensional group of
isometries along with the
 kinematic
quantities characterizing the fluid (see tables  11-14).
 Finally, in
section 5, we discuss the results obtained.

\section{Kinematic self-similarity}

In this section we present a brief summary of results on
kinematic self-similar vector fields and space-times admitting them.

\begin{enumerate}
\item The set of all KSS of the space-time forms a finite-dimensional
Lie algebra  under the usual Lie bracket operation
and will be denoted by $K_n$, where $n$ is its dimension. Furthermore,
it can be seen by direct computation that the Lie bracket of two 
arbitrary KSS is always a KV.

\item The set of all KSS with the same similarity index, 
$\kappa=\alpha/\delta$, forms also a 
finite-dimensional Lie algebra,  which will be denoted by
$K_s^{\kappa}$
(where $s$ is its dimension), and one has that $K_s^{\kappa}\subseteq K_n$
(i.e., $K_s^{\kappa}$ is a subalgebra of $K_n$).
Each non trivial $K_s^{\kappa}$ algebra, in the
sense that a proper-KSS (not a KV) exists, contains an
$(s-1)$-dimensional
subalgebra of KV. Equivalently, given two proper-KSS with the same 
similarity index $\kappa$, there always  exists a linear combination of them
which is a KV. The particular case $K_s^1$ (i.e., $\alpha=\delta$)
corresponds to the homothetic algebra.

\item Given two KSS,  $X_1$ and $X_2$, 
\bea
\L_{X_1} h_{ab}= 2 \delta_1 h_{ab}\ , & \quad & \L_{X_1} u_a=\alpha_1 u_a\ ,
\nonumber \\
\L_{X_2} h_{ab}= 2 \delta_2 h_{ab}\ , & \quad & \L_{X_2} u_a=\alpha_2 u_a\ ,
\nonumber
\eea
there always exist two vectors defined by
$$Y\equiv\alpha_2X_1-\alpha_1X_2\ ,\qquad Z\equiv\delta_2X_1-\delta_1X_2 \ ,$$
so that $Y$ is a KSS of type zero and $Z$ is a KSS of type infinite.

\item If two proper-KSS (not KV) of different similarity index exist in the
space-time, then any other KSS is a linear combination of them and the KVs.
From this consideration, it immediately follows  that if the dimension of the
isometric algebra of the space-time is $r$, then the maximum dimension 
of the kinematic self-similar algebra is $r+2$. In this case, one
can always construct two proper-KSS, such as one is of type zero and
the other is of type infinite. Furthermore, the space-time always admits an
$(r+1)$-dimensional homothetic algebra.

{\it Proof:}  Let $X_i\in K$, so that $\L_{X_i} h_{ab}= 2 \delta_i h_{ab}$,
$\L_{X_i} u_a=\alpha_i u_a$, $i=1,2,3$ and 
$\alpha_2\delta_1-\alpha_1\delta_2\not =0$, then 
$V\equiv \left( (\alpha_2-\delta_2)X_1 +(\delta_1-\alpha_1)X_2
\right)/(\alpha_2\delta_1-\alpha_1\delta_2)$ is a HVF, and
$W\equiv (\alpha_2\delta_1-\alpha_1\delta_2)X_3 +
(\alpha_3\delta_2-\alpha_2\delta_3)X_1 +
(\alpha_1\delta_3-\alpha_3\delta_1)X_2$ is a KV.

\item For a four-dimensional manifold, the highest possible dimension of
a kinematic self-similar algebra is $n=12$, and  can only occur
(as well as for  $n=11$) when the connection is flat and 
obviously  no perfect fluid solution exists.

The case $n=10$ is impossible, as it follows from considerations of the
dimension of the isometric algebra.

If $n=9$ then, the Killing subalgebra is 7-dimensional and multiply
transitive acting on the 4-dimensional manifold. The resulting
3-parameter Killing isotropy then implies that the Weyl tensor has Petrov
type $O$, and since the space-times must admit a homothetic algebra, 
known results regarding fixed points of the homothety \cite{Hall90}
implies that the Ricci tensor has Segre type $\{(2,11)\}$. It then 
follows that the space-time is a conformally flat homogeneous generalized
plane wave (the \lq\lq null-fluid" case) and so, there cannot be perfect
fluid solutions.

\end{enumerate}

\section{Analysis}

In this section, we study the different LRS space-times admitting a 
four-dimensional group of
 isometries, and
we provide  coordinate expressions for the metric and the
kinematic self-similar vector field.

Given the coordinate forms of the KVs for the metrics (\ref{r8})-(\ref{r10}),
 assuming the existence  of a KSS, $X$, since its 
commutator with a KV must be a KV, the Jacobi identities imply that, in these
coordinates, the
KSS  can only take the forms given in  table 1,

\[
\begin{array}{|c|c|c|}\hline
\mbox{case} & \mbox{ Killing vectors} & \mbox{KSS,}\ X \\ \hline\hline
\mbox{(i)}& \sin z \partial_y+{\displaystyle\Sigma'  \over \displaystyle\Sigma} \cos z
\partial_z & X^t(t) \partial_t +(\hat X^x(t)+nx )\partial_x +my \partial_y \\
 &
\cos z \partial_y-{\displaystyle\Sigma' \over \displaystyle\Sigma}  \sin z
\partial_z & m,n \in \R \\
 &\partial_x & m=0 \ {\rm for} \ k\not= 0 \\
 & \partial_z & \\ \hline

\mbox{(ii)} & \cos z \Delta \partial_x + \sin z \partial_y +  \cos z
{\displaystyle\Sigma' \over \displaystyle\Sigma} \partial_z
&  X^t(t) \partial_t +(\hat X^x(t)+2mx )\partial_x +my \partial_y \\
 & \sin z \Delta \partial_x - \cos z \partial_y + \sin z {\displaystyle\Sigma'
\over \displaystyle\Sigma} \partial_z &  m \in \R \\
 &\partial_x & m=0 \ {\rm if} \ k\not= 0 \\
 & \partial_z & \\ \hline
\mbox{(iii)} & \partial_x -y \partial_y -z\partial_z &  X^t(t) \partial_t +
X^x(t)\partial_x  \\
 & z\partial_y -y\partial_z & \\
 & \partial_y & \\
 & \partial_z & \\
\hline
\end{array}
\]
\begin{center}
{\bf Table 1.} {\small  KVs and  possible
KSS for the different
LRS metrics (i), (ii) and (iii).}
\end{center}

\noindent where a prime indicates a derivative with respect to $y$ and
\be
\Delta = \Gamma\left({\Gamma' \over \Gamma}-{\Sigma' \over
\Sigma}\right) \ . 
\ee

{}From now on, we will devote our study only to those space-times which can be
interpreted as perfect fluid solutions.
The
energy momentum tensor, $T_{ab}$, is  given by
\be
T_{ab}=(\mu+p)u_au_b+pg_{ab} \  ,
\ee
where $\mu$ is the energy density, $p$  the pressure, and $u^a$ the
four-velocity  of the fluid.

By simple inspection of the field equations, one can figure out which
components of the four-velocity vanish. Then,
solving 
 the kinematic self-similar conditions (\ref{k1})
 in the covariant unknowns $u_a$ and $h_{ab}$ for the contravariant $X$
 given in table 1,  we obtain restrictions for $X$ and, depending on the
 value of the parameters $\epsilon$, $\alpha$ and $\delta$, we obtain
 the different forms of the metric functions in each case.
 \hfill\break

\noindent{\bf Case (i)}

The possible forms of the four-velocity $u$ and the KSS $X$ for the line
element (\ref{r8})  are those given in
table 2,

\[
\begin{array}{|c||c|c|}\hline
\mbox{(i)} & u& X \\ \hline\hline
\epsilon =-1 & -dt & (\alpha t+ \beta) \partial_t +nx \partial_x +my
\partial_y \\ \hline
\epsilon =+1 & -A(t)dx &  (\delta t+ \beta) \partial_t +nx \partial_x +my
\partial_y \\ \hline \end{array}
\]
\begin{center}
{\bf Table 2.} {\small Four-velocity   and  KSS 
for the metric (\ref{r8}).}
\end{center}

\noindent where $\beta$ is a constant. 

For $\epsilon =-1$ and $k=0$,  the
space-time (\ref{r8}) always admits a proper-KSS, 
say  $\bar X = x \partial_x +y
\partial_y $,  of type zero 
 independently of the form of the metric functions.
It corresponds to the case $\alpha=\beta=0$. Apart from this case,
 the following possibilities arise

\[
\begin{array}{|c|c|c|c|}\hline
\epsilon =-1 & \mbox{case} & A(t) & B(t) \\ \hline\hline
\mbox{(i.a)} & \alpha \not=0,\ \beta =0 &t^{(\delta-n)/\alpha}& 
b\, t^{(\delta-m)/\alpha}  \\ \hline
\mbox{(i.b)} & \alpha=0,\ \beta  \not=0 &
 \exp\left({\delta-n\over \beta}t\right) & b\,\exp\left( {\delta-m\over
\beta}t\right) \\ \hline \end{array}
\]
\begin{center}
{\bf Table 3.} {\small Metric functions for the 
line element (\ref{r8}) with $\epsilon =-1$.}
\end{center}
where $b$ is a constant. In the case (i.a)
 since $\alpha \not=0$, by a translation of $t$,
 we can set $\beta$ to zero and, by rescaling $X$   with a
factor $1/\alpha$, we can set the parameter $\alpha$  to unity.  
In the case (i.b),
 we can also set $\beta$ to 1. 
Note that, for $k=0$, the case (i.a) admits two proper-KSS, 
$X$ and $\bar X$, of
different similarity index and therefore, the space-time is homothetic
 and, in the case (i.b), since $X$ and $\bar X$ are two
independent KSS of type zero,  the group
$G_4$ is not maximal.

For $\epsilon =+1$, the vector $\hat X= x\partial_x$ is a proper-KSS of 
type infinite. It
  satisfies $\L_{\hat X}u_a
=u_a$, $\L_{\hat X}h_{ab}= 0$  for all  $A(t)$
and $B(t)$. The remaining possibilities are given in table 4.

\[
\begin{array}{|c|c|c|c|}\hline
\epsilon =+1 & \mbox{case} & A(t) & B(t) \\ \hline\hline
\mbox{(i.c)} & \delta \not=0,\ \beta =0 &t^{(\alpha-n)/\delta}& 
b\, t^{(\delta-m)/\delta}  \\ \hline
\mbox{(i.d)} & \delta=0,\ \beta  \not=0 &
\exp\left({\alpha-n\over \beta}t\right) & b\,\exp\left(-{m\over \beta}t\right)
\\ \hline \end{array}
\]
\begin{center}
{\bf Table 4.} {\small Metric functions for the 
line element (\ref{r8}) with $\epsilon =+1$.}
\end{center}
Again, by normalizing $X$ we can set $\delta$ or $\beta$ to 1. Note that
 the case (i.c) admits 
also a proper-HVF, and in the case (i.d), 
 there exists a  further KV and $G_4$ is  not maximal.
\hfill\break

\noindent{\bf Case (ii)}

The four-velocity and the KSS for the line element (\ref{r9}) must have
the forms given in table 5.
\[
\begin{array}{|c||c|c|}\hline
\mbox{(ii)} & u& X \\ \hline\hline
\epsilon =-1 & -dt & (\alpha t+ \beta) \partial_t +2mx \partial_x +my
\partial_y \\ \hline
\epsilon =+1 & -A(t)\,(dx+\Gamma dz )&  (\delta t+ \beta) \partial_t + 2mx
\partial_x +my \partial_y \\ \hline \end{array}
\]

\begin{center}
{\bf Table 5.} {\small Four-velocity and KSS for the metric (\ref{r9}).}
\end{center}
The  only possible cases are those listed in tables 6 and 7,
 \hfill\break
\[
\begin{array}{|c|c|c|c|}\hline
\epsilon =-1 & \mbox{case} & A(t) & B(t) \\ \hline\hline
\mbox{(ii.a)} & \alpha \not=0,\ \beta =0 &a\, t^{(\delta-2m)/\alpha}& 
b\, t^{(\delta-m)/\alpha}  \\ \hline
\mbox{(ii.b)} & \alpha=0,\ \beta  \not=0 &
a\, \exp\left({\delta-2m\over \beta}t\right) & b\,\exp\left( {\delta-m\over
\beta}t\right) \\ \hline \end{array}
\]

\begin{center}
{\bf Table 6.} {Metric functions for the line element (\ref{r9}) with 
$\epsilon =-1$.}
\end{center}

\[
\begin{array}{|c|c|c|c|}\hline
\epsilon =+1 & \mbox{case} & A(t) & B(t) \\ \hline\hline
\mbox{(ii.c)} & \delta \not=0,\ \beta =0 &a\, t^{(\alpha-2m)/\delta}& 
b\, t^{(\delta-m)/\delta}  \\ \hline
\mbox{(ii.d)} & \delta=0,\ \beta  \not=0 &
a\,\exp\left({\alpha-2m\over \beta}t\right) & b\,\exp\left(-{m\over
\beta}t\right) \\ \hline \end{array}
\]

\begin{center}
{\bf Table 7.} {Metric functions for the line element (\ref{r9}) with 
$\epsilon =+1$.}
\end{center}
where  $a$ and $b$ are constants.
In each case, we can set $\alpha$,   $\beta$ or $\delta$ to
unity respectively.
 \hfill\break

\noindent{\bf Case (iii)}

By simple inspection of the field equations for the metric (\ref{r10}), 
we obtain 
\be
u=u_t(t)dt +u_x(t)dx \ , \label{k12}
\ee
subject to the condition
\be
u_t(t)^2= -\epsilon +{u_x(t)^2\over A(t)^2} \  . \label{k13}
\ee
The kinematic self-similar equations (\ref{k1}) 
specified to
the KSS  given in table 1 yield 
\be
u_{x,t}X^t=\alpha u_x
\ , \label{k14}
\ee
\be
u_{t,t}X^t=\epsilon (\delta-\alpha)u_t{{u_x}^2\over A^2}
\ , \label{k15}
\ee
\be
{X^t}_{,t}=\delta +\epsilon (\delta-\alpha) {u_t}^2
\ , \label{k16}
\ee
\be
{X^x}_{,t}=-2\epsilon (\delta-\alpha) {u_tu_x\over A^2}
\ , \label{k17}
\ee
\be
(A^2)_{,t}X^t= 2\delta A^2 -2\epsilon (\delta-\alpha) {u_x}^2
\ , \label{k18}
\ee
\be
(B^2)_{,t}X^t +2B^2X^x =2\delta B^2
\ . \label{k19}
\ee
We  study now two different cases:

\underline{\it  Comoving fluid flow}

In this case the component $G_{tx}$ of the Einstein tensor must vanish.
This yields
\be
A(t)=aB(t) \  , \quad  a\in \R \ ,
\label{k20}
\ee
and the following possibilities arise
\[
\begin{array}{|c||c|c|}\hline
\mbox{(iii)} & u& X \\ \hline\hline
\epsilon =-1 & -dt & (\alpha t+ \beta) \partial_t +n \partial_x\\ \hline
\epsilon =+1 & -A(t) dx &  (\delta t+ \beta) \partial_t + n
\partial_x \\ \hline \end{array}
\]

\begin{center}
{\bf Table 8.} {\small Four-velocity and KSS for the metric (\ref{r10})
with a comoving fluid flow.}
\end{center}

\noindent where  $\beta$ and $n$ are constants.
For  $\epsilon =-1$  the functional forms of the metric functions are given in
table 9. We  note that, in all cases, the constant $n$ must vanish, and 
the solutions are special cases of FRW models.
  \hfill\break
\[
\begin{array}{|c|c|c|c|}\hline
\epsilon =-1 & \mbox{case} & B(t) &\mbox{constraints} \\ \hline\hline
\mbox{(iii.a)} & \alpha \not=0,\ \beta =0 & t^{\delta/\alpha}& 
n=0  \\ \hline
\mbox{(iii.b)} & \alpha=0,\ \beta  \not=0 &
\exp\left({\delta\over \beta}t\right) &n=0\\ \hline \end{array}
\]
\begin{center}
{\bf Table 9.} {\small Metric functions for the line element (\ref{r10}) with
$\epsilon =-1$ in the comoving case.}
\end{center}
For $\epsilon =+1$ the metric functions are those given next in table 10.

\[
\begin{array}{|c|c|c|c|}\hline
\epsilon =+1 & \mbox{case} & B(t) &\mbox{constraints}  \\ \hline\hline
\mbox{(iii.c)} & \delta \not=0,\ \beta =0 &t^{(\delta-n)/\delta} &  \alpha=
\delta -n\\ \hline
\mbox{(iii.d)} & \delta=0,\ \beta  \not=0 &\exp\left(-{n\over
\beta}t\right) & \alpha=-n \\ \hline \end{array}
\]
\begin{center}
{\bf Table 10.} {\small Metric functions for the line element (\ref{r10}) with
$\epsilon =+1$ in the comoving case.}
\end{center}

\underline{\it Non-comoving fluid flow}

In this case $u_t\not=0$, $u_x\not=0$ and $A(t)\not=aB(t) $. We
distinguish two cases whereas $u_{t,t}$ vanish or not. The case $u_{t,t}=0$
can be fully integrated. Demanding $X$ not to be a KV, it yields
 \be
u=\beta dt+ctdx\ , \quad 
X= t\partial_t +n\partial_x \  , \quad\beta, c, n \in \R \,
\ee
\be
A(t)={ct\over \sqrt{\beta^2 +\epsilon }}\ , \quad B(t)=t^{1-n}\ , \label{k25} 
\ee
which is a homothetic space-time with $\alpha = \delta =1$. 

Let us consider now the case $u_{t,t}\not = 0$. From equations (\ref{k13}),
(\ref{k15}) and (\ref{k16}), we obtain
\be
-\alpha +2 \epsilon(\delta-\alpha){u_t}^2-\epsilon(\delta-\alpha) u_t({u_t}^2
+ \epsilon) {u_{t,tt}\over (u_{t,t})^2}=0 \ . \label{k26}
\ee
 Hence, it readily follows that proper homothetic solutions are not
possible. For $\alpha \not = \delta $, equation  (\ref{k26}) can be integrated
once and we get
\be
({u_t}^2)_{,t}=c({u_t}^2)^{\delta/2-\alpha \over
\delta-\alpha}({u_t}^2+\epsilon)^{\delta-\alpha/2 \over
\delta-\alpha}\ , \quad c\in\R \ .\label{k27}
\ee
Explicit solutions can be obtained  for some particular values of
the parameters $\alpha$ and $\delta$ but, in general, 
the solution of equation (\ref{k27}) will be given in an implicit
form. 
The remaining quantities can be obtained in terms of $u_t$ 
\be
X^t={2\epsilon\over c}(\delta-\alpha)({u_t}^2)^{\delta/2 \over
\delta-\alpha}({u_t}^2+\epsilon)^{-\alpha/2 \over
\delta-\alpha}\ ,
\ee
\be
u_x= e\left( {{u_t}^2\over {u_t}^2+\epsilon}
\right)^{{\alpha \over 2(\delta-\alpha)}}\ , \quad e\in\R \ ,
\ee
\be
A^2={{u_x}^2\over {u_t}^2+\epsilon} \ .
\ee
Then, integrating (\ref{k17}) we get $X^x$, and from (\ref{k19}) we 
obtain $B(t)$.

\section{Perfect fluid solutions}
In this section we  present  the different perfect fluid solutions admitting a
KSS with a maximal group  $G_4$ of isometries
with  an explicit mention of the physical quantities
characterizing the fluid; namely: density $\mu$, $\gamma$ 
(appearing in the equation of state $p=(\gamma-1)\mu$),
acceleration $\dot u_a$, vorticity $\omega_{ab}$, shear $\sigma$, 
volume expansion $\theta$, and deceleration parameter $q$.
In what follows, we have
eliminated some of the parameters, 
 in order to
simplify the expressions. \hfill\break

{\bf Case (i)}, $\epsilon =-1$.

The matter variables are
\be
\mu =  2{\dot A \dot B \over AB}+ {k\over B^2}+ \left( {\dot B \over B}
\right)^2\ , \quad
p=  -2{\ddot B \over B} -{k\over B^2}- \left( {\dot B \over B}
\right)^2 \ , \ee 
where a dot indicates a derivative with respect to $t$ and the non-trivial 
field equation is
\be
 {\ddot B\over B}+ {k\over B^2}+ \left( {\dot B \over B}
\right)^2 -{\dot A \dot B \over AB}- {\ddot A \over A}=0 \ .
\label{k28}
\ee
The perfect fluid solutions for the case (i.a) are given in table 11.

\begin{center}
\begin{tabular}{|c||c|c|c|c|c|c|c|c|} \hline
(i.a) & $\mu$ & $\gamma$ & $\delta$ & $m$
& $b^2$ & $\sigma^2$& $\theta$ & $q$ \\ \hline \hline
$k=1$ & ${\displaystyle n^2-4n+3 \over\displaystyle t^2}$ & ${\displaystyle
2 \over\displaystyle 3-n}$ & $1$ & $0$ & ${\displaystyle 1\over\displaystyle
n^2-2n}$ &$ {\displaystyle n^2 \over\displaystyle 3t^2}$& 
${\displaystyle 3-n \over\displaystyle t}$ & $ {\displaystyle n \over
\displaystyle  3-n}$ \\ \hline

$k=0$ & ${\displaystyle n(4-3n) \over\displaystyle 2t^2}$ & $2$ & $1$
 & ${\displaystyle 2-n \over
\displaystyle 2}$ & $1$ & ${\displaystyle (2-3n)^2 \over\displaystyle 12t^2}$
 & ${\displaystyle 1 \over\displaystyle t}$ & $2$
 \\ \hline

$k=-1$ & ${\displaystyle n^2-4n+3 \over\displaystyle t^2}$ & ${\displaystyle
2 \over\displaystyle 3-n}$ & $1$ & $0$ & ${\displaystyle
1\over\displaystyle 2n-n^2}$ & ${\displaystyle n^2 \over\displaystyle 3t^2}$
 & ${\displaystyle 3-n \over\displaystyle t}$ &
$ {\displaystyle n \over\displaystyle 3-n}$
\\ \hline  \end{tabular}
\end{center}

\begin{center}
{\bf Table 11} {\small Perfect fluid solutions for the case (i.a).}
\end{center}

\noindent For all the above cases we have $\dot
u_a=\omega_{ab} =0$
(geodesic and irrotational flow) and $X$ is a proper-HVF. All of them are special
cases of Szekeres-Szafron universes \cite{Szafron}. There exists another
possible solution
for $k=0$ and $n=m$, but this is a special type of FRW model and, in
keeping with the
assumption of maximality of $G_4$, we have not listed it here. Also,
notice
that the case $k=0$, for the particular value  $m={2 \over 3}$, corresponds
to an FRW
model as well.

For the case (i.b) there is no solution for $k=-1$; for $k=1$, $X$ becomes a KV;
and for $k=0$,  the group $G_4$ is not maximal. 

The only possible proper kinematic self-similar solution (not homothetic),
for the metric (\ref{r8}) with $\epsilon=-1$,
corresponds to $k=0$ without any further assumption a priori for  the metric
functions. Thus,  integrating equation (\ref{k28}) for $k=0$, we obtain
\be
A(t)=c_1B(t) +c_2 B(t) \int {dt \over B(t)^3} \  ,
\ee
where $c_1$ and $c_2$ are constants and B(t) is an
 arbitrary function.
For an equation of state $p=(\gamma -1)\mu$, the solution for $\gamma\not =2$
 becomes
\cite{Stewart68}
$$A= (B^{3(2-\gamma)/2}+c)^{1/(2-\gamma)} B^{-1/2} \ ,$$
\be
t=\int (B^{3(2-\gamma)/2}+c)^{(\gamma-1)/(2-\gamma)}B^{1/2} dB \ ,
\ee
$$\mu={3\over (AB^2)^{\gamma}} \ , $$
where $c$ is a constant, and for   $\gamma =2$ it corresponds to
the  homothetic solution listed in {\bf Table 11}. \hfill\break

{\bf Case (i)}, $\epsilon =+1$.

The field equations reduce to
\be
\mu =   -2{\ddot B \over B} +{k\over B^2}- \left( {\dot B \over B}
\right)^2 \ , \quad
p=  2{\dot A \dot B \over AB}- {k\over B^2}+ \left( {\dot B \over B}
\right)^2\ ,  \ee
and
\be
 {\ddot B\over B}+ {k\over B^2}-\left( {\dot B \over B}
\right)^2 -{\dot A \dot B \over AB}+ {\ddot A \over A}=0 \ ,
\label{k29}
\ee
Since $\hat X=x\partial_x$ is always a proper-KSS,
 any pair of functions $A(t)$ and $B(t)$ satisfying (\ref{k29}) will
represent a kinematic self-similar solution if they satisfy the energy
conditions. The only  solutions  corresponding to homothetic
space-times are listed in table 12.

\begin{center}
\begin{tabular}{|c||c|c|c|c|c|} \hline
(i.c) & $\mu$ & $\gamma$  & $\alpha$ & $m$ & $b^2$ \\ \hline \hline
$k=1$ & ${\displaystyle 1-n^2 \over\displaystyle t^2}$ & ${\displaystyle 2
\over\displaystyle 1+n}$ & $1$ & $0$ & ${\displaystyle
1\over\displaystyle 2-n^2}$  \\ \hline $k=0$ & ${\displaystyle n(1-n^2)(3n-4)
\over \displaystyle (2-n)^2t^2}$ & ${\displaystyle 2 \over\displaystyle 
1+n}$  & $1$ &
${\displaystyle 2-n^2 \over\displaystyle 2-n}$ & $1$  \\ \hline \end{tabular}
\end{center}

\begin{center}
{\bf Table 12} {\small Perfect fluid solutions for the case (i.c).}
\end{center}

\noindent Note that $t$ here is a spacelike coordinate and all
these solutions are
stationary. They
have $\sigma = \theta = \omega_{ab} =0 $,
whereas $\dot u_a={1-n \over t} \delta^t_a$.
 The case $k=1$ is further discussed in 
\cite{Ali2,Haggag,Ibanez82} 
corresponding to a static spherically symmetric solution.
 No physically
realistic solutions exist for $k=-1$ since the energy density is then
 negative necessarily. 
\hfill\break

{\bf Case (ii)}, $\epsilon =-1$.

The only possible  kinematic self-similar solution with a maximal
group $G_4$ is given in table 13.

\begin{center}
\begin{tabular}{|c||c|c|c|c|c|} \hline
(ii.a) & $\mu$ & $\gamma$ & $\delta$ &
$a^2b^{-4}$ \\ \hline \hline
$k=0$ & ${\displaystyle 6-17m+12m^2 \over \displaystyle 2t^2}$ &  
${\displaystyle 4-6m \over\displaystyle 6-17m+12m^2}$ & $1$   & ${\displaystyle
m \over \displaystyle 2}-m^2$ \\  \hline \end{tabular} 
\end{center}
 
\begin{center}
{\bf Table 13} {\small Perfect fluid solution for the case (ii.a).}
\end{center}

\noindent It corresponds to a homothetic solution. The  fluid is 
geodesic, irrotational and
\be
\sigma^2={m^2\over 3t^2} \ , \quad \theta={3-4m \over t} \ , \quad  q={4m
\over 3-4m} \ . 
\ee
The case $k=1$ corresponds again to a special
type of FRW model with equation of state $\mu + 3p=0$, whereas no physically
significant solutions exist for $k=-1$, since the energy density is always
negative. 

Case (ii.b) is empty of solutions with a maximal group $G_4$. The case $k=1$
corresponds to an FRW model and there are no perfect fluid 
solutions in the other cases.

\hfill\break

{\bf Case (ii)}, $\epsilon =+1$.

In table 14, we give all perfect fluid solutions.

\begin{center}
\begin{tabular}{|c||c|c|c|c|} \hline
(ii.c) & $\mu$  & $\gamma$ & $\alpha$ &$a^2$\\\hline\hline
$k=1$ & ${ 2-{1 \over 2b^2}\over\displaystyle t^2}$ &  $2$ & $1$
& $2b^2(2b^2-1)$ \\ \hline 
$k=0$ & ${\displaystyle 2+m-6m^2 \over\displaystyle t^2}$  & ${\displaystyle
4-6m \over \displaystyle 2+m-6m^2}$ & $1$   &$(1-m-m^2)b^4$ \\ 
\hline $k=-1$ & ${ 2+{1 \over 2b^2}\over \displaystyle t^2}$ & 
$2$& $1$  & $2b^2(2b^2+1)$\\\hline  \end{tabular}
\end{center}

\begin{center}
{\bf Table 14} {\small Perfect fluid solutions for the case (ii.c).}
\end{center}

\noindent Note that all of them 
 are  stationary and homothetic with
\be
\dot u_a= {1-2m \over t}
\delta_a^t \ , \quad
\sigma = \theta =0 \ .
\ee
The only
non-vanishing component of the vorticity tensor is 
\be\omega_{yz}={a\over
2}t^{1-2m}\Gamma' \ .
\ee

For the 
case (ii.d) there are no solutions for $k=1$ and $k=0$, and for $k=-1$, $X$
is a KV therefore, the $G_4$ is  not maximal.

\hfill\break

{\bf Case (iii)}

\underline{\it Comoving fluid flow.}

The only possible solutions of Einstein's field equations for a
perfect fluid correspond in this case to
 $\epsilon =-1$, case (iii.a), the solution being again a  special type of
  FRW model with an equation of state $\mu+3p=0$.
 For $\epsilon =+1$, (iii.c) corresponds to 
a homothetic vacuum space-time, and there is no solution for the case (iii.d).

\underline{\it Non-comoving fluid flow.}

The Einstein's field equations
for the  case $u_{t,t}=0$ (i.e., solution (\ref{k25}))
imply $\beta^2+\epsilon = c^2(1-n)^2$ and hence, $\mu +p=0$. Therefore,
no perfect fluid exits with $\mu +p >0$.

The case in which $u_{t,t}\not=0$,  homothetic solutions are not possible.
Furthermore, given the solution of equation 
 (\ref{k27}) one has to check the existence or not of perfect fluid 
 solutions for the different value of the parameters
$\alpha$ and $\delta$.

\section{Discussion}
To summarize, we have reviewed the concept of kinematic self-similar
vector fields, presented some results on space-times admitting them, and
we have
studied perfect fluid solutions which admit a maximal group $G_4$ of
isometries  acting multiply transitively, together with a kinematic
self-similarity.

In this case, the solutions are LRS and the maximal group $G_4$ acts
on three-dimensional non-null orbits. Their metrics are given by the
equations (\ref{r8}) to (\ref{r10}). With the exception of (\ref{r8})
with $k=+1$, that does not admit a simply transitive subgroup $G_3$
of isometries, these metrics
are all special cases of the Bianchi models (see table 11.1 in \cite{Kramer}
for subgroups $G_3$ on $V_3$ occurring in solutions with 
multiply-transitive groups). The  possible $G_3$ are:
for (\ref{r8}), $k=0$, $G_3I$ or $G_3VII_0$; 
for (\ref{r8}), $k=-1$, $G_3III$; 
for  (\ref{r9}), $k=+1$, $G_3IX$; 
for  (\ref{r9}), $k=0$,  $G_3II$;
for (\ref{r9}), $k=-1$, $G_3VIII$ or  $G_3III$; 
and for (\ref{r10}), $G_3V$ or  $G_3VII_h$
(see \cite{Kramer} chapters 11, 12 and 29 for a review of perfect
fluid solutions). 

We have examined the canonical line-elements of the space-times 
admitting a maximal group $G_4$ acting on three-dimensional non-null
orbits, one by one, for the possible admission  of the additional
symmetry and then, we have restricted  the source to be a perfect
fluid with an arbitrary equation of state.

Many cases are empty of solutions with a maximal group $G_4$, and
when  they exist, the proper-KSS vector usually becomes a homothetic
vector and hence, the perfect fluid satisfies a linear equation
of state (i.e., $p=(\gamma-1)\mu$, where $\gamma$ is a constant).

In this paper, we explicitely give {\it all} the homothetic solutions
admitting a  maximal group $G_4$ of isometries (see tables 11-14) 
along with the kinematical quantities characterizing the fluid.
We realize that for the case (ii.c) the perfect fluid solutions have
a non-vanishing component of the vorticity tensor. This result rejects
some speculations that kinematic self-similar  models were vorticity free.
We notice that in this particular case, the isometry algebra acts on
three-dimensional timelike orbits and therefore, the solutions are
stationary.

Non-homothetic kinematic self-similar solutions are only possible for
cases (i) and (iii). For case (i), the orbits associated with the 
kinematic self-similar and isometric algebra, respectively,
 must necessarily coincide.
For the metric (i) with $\epsilon =-1$, the only non-homothetic solution
corresponds to $k=0$ (i.e., plane symmetry). The solution is given by
equation (26). In this case, the fluid satisfies a barotropic  equation
of state that is not necessarily linear. For a linear equation of 
state, with $\gamma \not= 2$, the solution (27) was found previously
by Stewart and Ellis
\cite{Stewart68}, and the stiff case ($p=\mu$) corresponds to a homothetic
solution. For the metric (i) with  $\epsilon =+1$, any pair of functions
$A(t)$ and $B(t)$ satisfying equation (29) will represent a 
kinematic self-similar solution.

In case (iii), the kinematic self-similar orbits are four-dimensional
and the fluid flow is restricted to be non-comoving.

\section*{Acknowledgements}
The author would like to thank the referees for their comments.
Financial support from
 the European Union, 
TMR Contract No. ERBFMBICT972771 is also acknowledged.

\end{document}